\documentclass[pra,reprint]{revtex4-1}
\usepackage{graphicx,url}
\usepackage{amsmath,braket}

\renewcommand{\vec}[1]{\mathbf{#1}}

\begin{document}
\title{Quantum scar affecting
  the motion of three interacting particles in a circular trap}
\author{D.J.\ Papoular}
\email[Electronic address: ]{david.papoular@cyu.fr}
\author{B.\ Zumer}
\affiliation{LPTM, UMR 8089 CNRS \& CY Cergy Paris Universit\'e, Cergy--Pontoise, France}
\date{\today}

\begin{abstract}
We theoretically propose a quantum scar affecting the motion
of three interacting particles in a circular trap.
We numerically calculate the quantum eigenstates of the system and
show that some of them are scarred by a
classically unstable periodic trajectory, in the vicinity of which
the classical analog exhibits chaos.
The few--body scar we consider is
stabilized by quantum mechanics, and we analyze it along the lines of
the original quantum scarring mechanism
[Heller,  Phys.\ Rev.\  Lett.\ \textbf{53},  1515 (1984)].
In particular, we identify towers of
scarred quantum states which we fully explain
in terms of the unstable classical trajectory
underlying the scar.
Our proposal is within experimental reach owing to very recent advances
in Rydberg atom trapping.
\end{abstract}

\maketitle

\section{Introduction}

\begin{figure}
  \includegraphics[width=\linewidth]{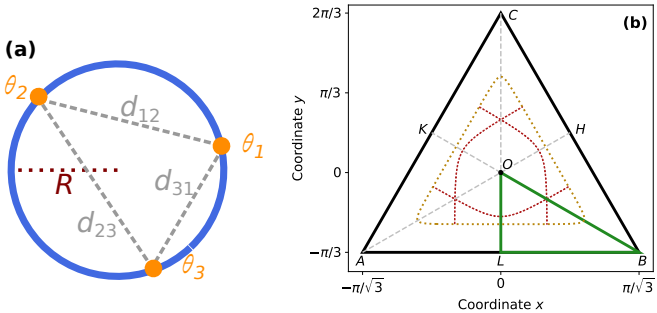}
  \caption{
    \label{fig:circle_triangle}
    \textit{(a)} Three particles (orange disks) interacting via a repulsive
    van der Waals interaction
    of strength $C_6>0$, constrained to move on a circle
    of radius $R$, their angular coordinates $\theta_i$
    and distances $d_{ij}$.
    \textit{(b)} The $(x,y)$ configuration space is the inside of the
    triangle defined by the points
    $A(-\pi/\sqrt{3},-\pi/3)$, $B(+\pi/\sqrt{3},-\pi/3)$,
    $C(0,2\pi/3)$.
    The dashed
    golden line limits the classically accessible region
    for the energy $E=7C_6/R^6$. The three dotted red lines show the three
    classical periodic
    trajectories of type $B$ for this energy.
    The small green triangle $OLB$ is the reduced configuration space
    within which quantum wavefunctions are calculated.
  }
\end{figure}

The thermalization of closed
interacting quantum systems \cite{polkovnikov:RMP2011}
may be impeded by various mechanisms
\cite{sutherland:WorldScientific2004,abanin:RMP2019}
whose investigation is strongly motivated by
contemporary applications \cite{gross:Science2017,zhu:PRL2022}.
Indeed, slowly--thermalizing systems retain memory of their initial state over longer times
\cite{alet:CRPhys2018}, making them useful for
quantum simulation \cite{gross:Science2017} and
quantum information processing \cite{zhu:PRL2022}.
Atomic systems are an excellent test--bed for chaos
\cite{blumel:CUP1997,friedrich:PhysRep1989,courtney:PRA1995},
and techniques for the individual manipulation \cite{browaeys:NatPhys2020}
of Rydberg atoms \cite{saffman:RMP2010} have extended its exploration to
interacting systems. A recent experiment on Rydberg atom arrays
\cite{bernien:Nature2017} has initiated the investigation of weak ergodicity breaking
in many--body systems
\cite{turner:NatPhys2018,serbyn:NatPhys2021}. Systems exhibiting this phenomenon
thermalize rapidly for most initial conditions, but specific initial states yield
non--ergodic dynamics. This behavior is analogous to the quantum scars
initially predicted \cite{heller:PRL1984} 
and observed \cite{stein:PRL1982}
in the absence of interactions, which also lead to weak ergodicity
breaking \cite{berry:ProcRSocLondA1989} by impacting some
\cite[chap.~22]{heller:Princeton2018} quantum eigenstates.
Hence, it is also called `many--body scarring' \cite{jepsen:NatPhys2022}.
A similar phenomenon has  been predicted in the context
of the Dicke model \cite{furuya:AnnPhys1992},
where the quantum scars are due to the collective light--matter interaction
and impact many
quantum eigenstates \cite{pilatowsky-cameo:NatCommun2021}.

Despite the intense theoretical scrutiny \cite{moudgalya:RepProgPhys2022},
only two experiments \cite{bernien:Nature2017,su:arXiv2022}
and one explicit proposal \cite{jepsen:NatPhys2022}
explore
many--body scarring so far
\cite{bernien:Nature2017,jepsen:NatPhys2022,su:arXiv2022}. In all three cases,
the observed non--ergodic behavior is linked to classical physics.
The experiments of Refs. \cite{bernien:Nature2017,su:arXiv2022}
both probe the PXP model \cite{ho:PRL2019} in regimes where the classical analog
system \cite{turner:PRX2021}
explores the vicinity of classically stable periodic trajectories,
so that the absence of thermalization may be traced back to the classical
Kolmogorov--Arnold--Moser theorem \cite[Sec.~VI]{michailidis:PRX2020}.
The proposal of Ref.~\cite{jepsen:NatPhys2022} refers to spin helices in
various geometries. Their classical limit is stable \cite{jepsen:PRX2021},
and from the quantum point of view they generalize
helices predicted \cite{popkov:PRB2021}
and observed \cite{jepsen:NatPhys2022} in the integrable XXZ chain.
Hence, the proximity of integrable models is expected to play a key role.

In this article, we propose a three--body system
hosting a quantum scar which relies on the interaction between particles.
It may be realized experimentally 
owing to very recent advances in Rydberg atom trapping
\cite{barredo:PRL2020,cortinas:PRL2020}.
It is simple enough to be fully analyzed by combining the 
numerical calculation of stationary states
and well--established tools for the analysis of chaotic systems
\cite{gutzwiller:Springer1990}, in the spirit of Heller's original
proposal \cite{heller:PRL1984}.

The system we consider exhibits ``towers'' of scarred states
which  are approximately  evenly spaced  in  energy. These  are a  key
feature of both quantum scars \cite{heller:LesHouches1989} and
many--body scars
\cite{turner:NatPhys2018,su:arXiv2022,moudgalya:RepProgPhys2022,
  choi:PRL2019,serbyn:NatPhys2021}.
In  the  present context, we explain them  in terms of the  classically unstable
periodic trajectory causing the scar, in the spirit of Heller's original
argument \cite[Fig.~22]{heller:LesHouches1989}.
The phase space dimensionality of the few--body system we consider (4, see below) matches the
maximum number of independent parameters introduced so far in the variational
approaches applied to the many--body PXP model and its generalizations
\cite[\S III.A]{michailidis:PRX2020}. In stark contrast to the many--body
PXP model where approximate classical limits have to be cleverly
constructed \cite{turner:PRX2021}, our few--body system affords an exact
reduction to four parameters  and the identification of the
classical analog is straightforward.

We formulate our proposal in terms of trapped Rydberg atoms
\cite{barredo:PRL2020,cortinas:PRL2020}.
However, we expect other interacting systems
with the same symmetries
to exhibit similar quantum scars. We substantiate this claim
in the Appendix (Sec.~\ref{sec:SM:HenonHeiles})
by identifying the quantum scar
for the H\'enon--Heiles (HH)
potential \cite{henon:AstronJ1064}.
In particular, the scar may be probed using three dipolar particles
\cite{baranov:ChemRev2012}.

\section{The considered system}

\begin{figure}
  \includegraphics[width=\linewidth]
  {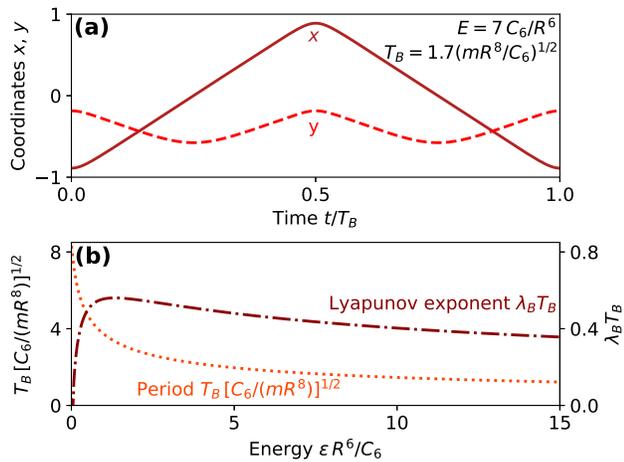}
  \caption{\label{fig:Rydberg_TrajB}
    \textit{(a)} Periodic trajectory $B$ for the energy $\epsilon=7\, C_6/R^6$,
    described by its coordinates $x(t)$ and $y(t)$
    as a function of time $t$.
    \textit{(b)} The period $T_B(\epsilon)$, and the product
    $\lambda_B\times T_B$ of the Lyapunov exponent and the period, for
    the periodic trajectory $B$ as a function of the energy $\epsilon$.
  }
\end{figure}

We consider three identical bosonic particles of  mass $m$ in a circular
trap of radius $R$. The Hamiltonian reads:
\begin{equation}
  H=(l_{1}^2+l_{2}^2+l_{3}^2)/(2mR^2)
  +v(d_{12})+v(d_{23})+v(d_{31})\ ,
\end{equation}
where $l_i$ is the component of the angular momentum of particle $i$
along the rotation axis, which is perpendicular to
Fig.~\ref{fig:circle_triangle}a. We assume that 
the interaction $v(d_{ij})$ between the particles $i$ and $j$   
only depends on their distance $d_{ij}=2R|\sin[(\theta_i-\theta_j)/2]|$.
For circular Rydberg atoms
whose electronic angular momenta are 
perpendicular to the plane, $v(d_{ij})=C_6/d_{ij}^6$ with $C_6>0$
\cite[App.~A]{nguyen:PRX2018}.

\begin{figure}
  \includegraphics[width=\linewidth]
  {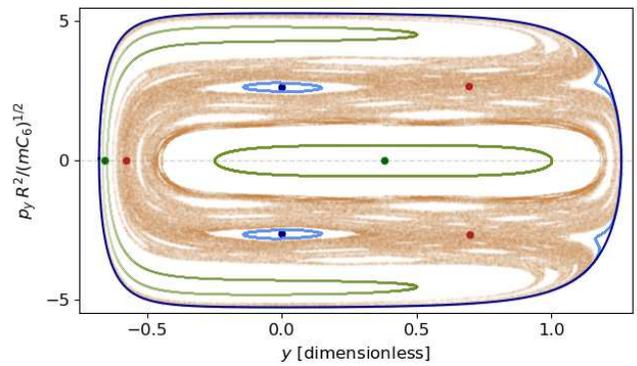}
  \caption{\label{fig:SoS}
    Classical surface of section \cite[\S 1.2]{lichtenberg:Springer1992}
    for the Hamiltonian of Eq.~(\ref{eq:Ham_Rydberg})
    with $p_z=0$,
    $\epsilon=7C_6/R^6$, $x=0$, and $p_x\geq 0$. The dark blue
    dots and outer curve indicate the periodic trajectories of type $A$;
    the red and green dots
    show those of types $B$, and $C$. The closed blue and green
    curves show non--ergodic trajectories near $A$ and $C$. The $\sim 287,000$
    thin brown dots all belong to the same ergodic trajectory.
    The periodic trajectories of type $B$, which yield the quantum scar,
    are all within
    the classically ergodic region.}
\end{figure}

We introduce the
Jacobi coordinates \cite[\S 1.2.2]{faddeev:Springer1993}
$x=[(\theta_1+\theta_2)/2-\theta_3+\pi]/\sqrt{3}$, 
$y=(\theta_2-\theta_1)/2 -\pi/3$,
$z=(\theta_1+\theta_2+\theta_3)/3 -2\pi/3$,
and their conjugate momenta $p_x$, $p_y$, $p_z$
(which carry the unit of action).
In terms of these,
$H=p_z^2/(3mR^2)+H_\mathrm{2D}$,
where
\begin{equation}
  \label{eq:Ham_Rydberg}
  H_\mathrm{2D}=(p_x^2+p_y^2)/(4mR^2)+V(x,y)
  \ .
\end{equation}
Here, $V(x,y)=v(x,y) C_6/R^6$, with
\begin{multline}
  \label{eq:pot_Rydberg}
  v(x,y)=[
  \sin^{-6}(\pi/3+y)+\sin^{-6}(\pi/3+x\,\sqrt{3}/2-y/2)\\
  +\sin^{-6}(\pi/3-x\,\sqrt{3}/2-y/2)]/64-1/9
  \ ,
\end{multline}
energies being measured from the  minimum $V(\vec{0})$.
The free motion of the coordinate $z$
reflects the conservation of the total angular
momentum $p_z=l_1+l_2+l_3$.
The Hamiltonian $H$ is invariant
\footnote{The full plane group characterizing the symmetries of
  $v(x,y)$ is $p6mm$ \cite[Part 6]{hahn:Springer2005}.}
under the point
group
$C_{3v}$
\cite[\S 93]{landau3:ButterworthHeinemann1977},
generated by the 3--fold rotation about the axis $(x=y=0)$
and the reflection
in the plane $(x=0)$.

\begin{figure*}
  \includegraphics[width=\linewidth]
  {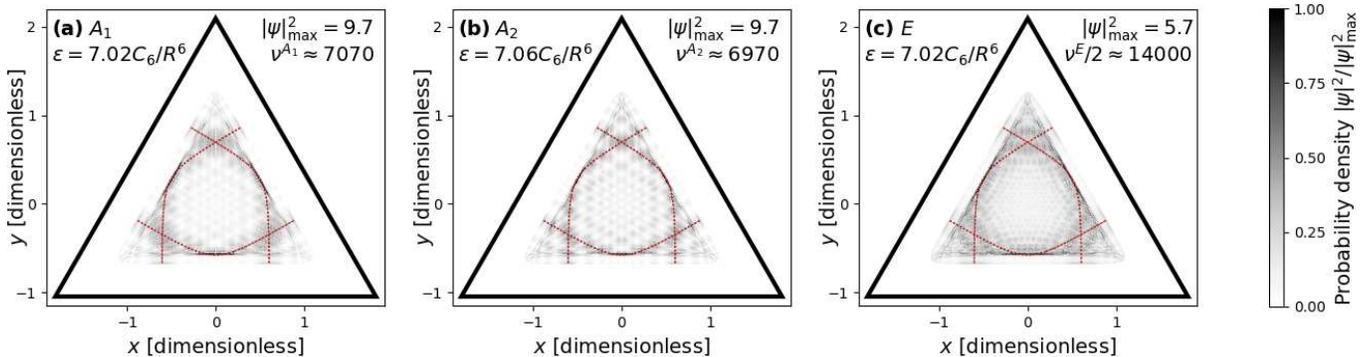}
  \caption{\label{fig:Rydberg_scars}
    Probability density $|\psi(x,y)|^2$ of the scarred quantum
    eigenstate whose energy
    is closest to $7C_6/R^6$ in each irreducible representation
    $\rho=$ \textit{(a)} $A_1$,
    \textit{(b)} $A_2$,
    and \textit{(c)} $E$.
    The dashed red lines show the
    three classically unstable periodic
    trajectories of type $B$ for the corresponding energy $\epsilon$.
    The  densities are maximal near the unstable trajectories,
    signalling the quantum scar.
    The integer $\nu^{(\rho)}$ in an approximation to the index of the
    shown quantum state  in the representation $\rho$.
    }
\end{figure*}

\section{Classical physics}
We first analyze the classical dynamics described by
the Hamiltonian $H$.
Expressing momenta, energies, and times  
in units of
$P_\mathrm{ref}=(mC_6/R^4)^{1/2}$,
$C_6/R^6$, 
and $(mR^8/C_6)^{1/2}$, respectively,
the classical results are independent of
$m$, $C_6$, and $R$,
leading to the scaled predictions in
Figs.~\ref{fig:circle_triangle}--\ref{fig:Rydberg_scars}.
We choose the rotating reference frame  such that $p_z=0$ and $z=0$.
The divergence of $v(d_{ij})$
prevents the particles from crossing, so that we assume
$\theta_1<\theta_2<\theta_3<\theta_1+2\pi$ at all times.
Hence,
the classical problem is reduced to a point moving in the 2D plane $(x,y)$
within the equilateral triangle of Fig.~\ref{fig:circle_triangle}b,
in the presence
of the potential $V(x,y)$.

We have characterized the periodic trajectories
of $V$
using our own C++ implementation of the numerical approach of
Ref.~\cite{baranger:AnnPhys1988}.
We find three families of periodic trajectories, existing 
for all energies $\epsilon>0$:
we label them $A$, $B$, $C$ in analogy with the
results for the HH potential \cite{davies:Chaos1992}.
We shall analyze them and their bifurcations
in a forthcoming paper \cite{papoular:InPrep2022}. Here,
we focus on family $B$, which yields the quantum scar. 
For a given $\epsilon$,
there are three trajectories of type $B$,
due to the three--fold rotational symmetry
of the potential $V$.
They are represented in the $(x,y)$ plane
in Fig.~\ref{fig:circle_triangle}b, and the one which is symmetric about the vertical
axis is shown as a function of time
in Fig.~\ref{fig:Rydberg_TrajB}a.
They are
unstable for all energies,
as shown by the Lyapunov exponent $\lambda_B>0$ in
Fig.~\ref{fig:Rydberg_TrajB}b.
Figure \ref{fig:Rydberg_TrajB}b shows that
trajectory $B$ satisfies both conditions heralding a quantum scar:
$\lambda_B T_B < 2\pi$
\cite[ch.~22]{heller:Princeton2018}, and lower values of $\lambda_B T_B$ signal
stronger scarring \cite[\S 9.3]{ozorioDeAlmeida:CUP1988}.
The unstable trajectory $B$ does not bifurcate
\cite[\S 2.5]{ozorioDeAlmeida:CUP1988}, so that the scar strengths
associated with it for all $E>0$ do not benefit from the classical enhancement due to
the proximity of bifurcations \cite{keating:ProcRSocLondA2001}.
This sets it apart from a  previous proposal  involving a
scar hinging on this enhancement \cite{prado:EPL2009} so that, in
stark contrast to ours, it is
captured by Einstein--Brillouin--Keller quantization
\cite{zembekov:JChemPhys1997}.

In order to visualize effects beyond 
the linear regime, Fig.~\ref{fig:SoS} shows the surface of section
\cite[\S 1.2]{lichtenberg:Springer1992}
of  $H$ for $\epsilon=7C_6/R^6$
and the conditions $x=0$, $p_x>0$ (allowing for
a comparison with the HH potential \cite{gustavson:AstronJ1966}). It
exhibits both non--ergodic regions comprising  tori
\cite[App.~8]{arnold:Springer1989}
and an ergodic region, as is typical for a non--integrable system
\cite[\S 1]{bohigas:PhysRep1993}.
The three fixed points corresponding 
to trajectories $B$ are all located in the ergodic region.
This precludes their stabilization by any classical mechanism.

\section{Quantum physics}
We seek the eigenfunctions of $H$ in the form
$\Psi_n(\theta_1,\theta_2,\theta_3)=\psi_n(\vec{r}) e^{in z}$,
where $\vec{r}=(x,y)$ and $n=p_z/\hbar$.
The wavefunction $\psi_n$ is an eigenstate of
$H_\mathrm{2D}$ with the energy $\epsilon$.
It is defined on the whole $(x,y)$
plane. Its symmetries are related to
\textit{(i)} angular periodicity,
\textit{(ii)} bosonic symmetry, and
\textit{(iii)} the point group $C_{3v}$.

We first discuss ${(i)}$ and ${(ii)}$.
\textit{(i)}
The $2\pi$--periodicity of $\Psi_n$ in terms of 
$(\theta_i)_{1\leq i\leq 3}$
yields $\psi_n(\vec{r}-\vec{BC})=\psi_n(\vec{r}-\vec{CA})=\psi_n(\vec{r}-\vec{AB})
=\psi_n(\vec{r})e^{-i2\pi n/3}$, so that $n$ is an integer.
\textit{(ii)}
The bosonic symmetry of $\Psi_n$
leads to
$\psi_n(\mathcal{S}\vec{r})=+\psi_n(\vec{r})$,
where $\mathcal{S}$ is the symmetry about
any of the lines $(AB)$, $(BC)$ or $(CA)$ in the $(x,y)$ plane.
Hence, we may restrict the configuration space to the inside of the
triangle $ABC$ of Fig.~\ref{fig:circle_triangle}b.
Along its edges,
$v(x,y)$ strongly diverges
(e.g.\ $v\approx(y+\pi/3)^{-6}$ near $[AB]$),
so that $\psi_n=0$ there.
Combining $\textit{(i)}$ and $\textit{(ii)}$,
and calling $\mathcal{R}$ the rotation
of angle $2\pi/3$ about $O$,
$\psi_n(\mathcal{R}\vec{r})=\psi_n(\vec{r})e^{2in\pi/3}$.

We now analyze the role of the point group $C_{3v}$.
We classify the energy levels in terms of its three irreducible
representations $\rho=A_1$, $A_2$, and $E$
\cite[\S 95]{landau3:ButterworthHeinemann1977}. Hence, Hilbert
space is split into three unconnected blocks. These may be told
apart through the behavior of $\psi_n$ under
two operations in the $(x,y)$ plane \cite{feit:JComputPhys1982}:
$\mathcal{R}$ and the
reflection $\mathcal{S}_\Delta$ about the line $\Delta=(CL)$ 
(see Fig.~\ref{fig:circle_triangle}b). Wavefunctions pertaining to the
1D representations $A_1$
or $A_2$ satisfy $\psi_n(\mathcal{R}\vec{r})=\psi_n(\vec{r})$,
so that $n=0$ modulo 3. Under reflection,
$\psi_n(\mathcal{S}_\Delta\vec{r})=\pm\psi_n$,
where the $+$ and $-$ signs hold for
$A_1$ and $A_2$, respectively.
Wavefunctions pertaining to the 2D representation $E$
satisfy $\psi_n(\mathcal{R}\vec{r})=\exp(\pm 2i\pi/3)\psi_n(\vec{r})$,
so that $n= \pm 1$ modulo 3 \cite{brack:Chaos1995}.
Then, exploiting time--reversal invariance
(see Sec.~\ref{sec:irrep_E} in the Appendix)
we may choose the two degenerate basis states
to be $\psi_n$ and its complex conjugate
$\psi_n^*$ with $\psi_n(\mathcal{S}_\Delta\vec{r})=\psi_n^*(\vec{r})$.

These symmetry considerations further reduce the configuration space
to the green triangle $OLB$ of Fig.~\ref{fig:circle_triangle}b.
We deal with representations $A_1$, $A_2$, and $E$
separately by applying different boundary conditions
on its edges
(see Sec.~\ref{sec:SM:boundaryconditions} in the Appendix).
We solve the resulting stationary Schr\"odinger equations
using the finite--element software FreeFEM \cite{hecht:JNumerMath2012}.
The classical scaling  no longer holds.
Instead, the energy spectra and wavefunctions depend on the
dimensionless ratio $\eta=\hbar/P_\mathrm{ref}=\hbar R^2/(mC_6)^{1/2}$.
Smaller values of $\eta$ signal 
deeper quasiclassical behavior:
we choose $\eta=0.01$.
We focus on energies $\epsilon\sim 7C_6/R^6$, which
are large enough for the classical
ergodic trajectory (brown dots on Fig.~\ref{fig:SoS})
to occupy a substantial part of phase space.

Figure \ref{fig:Rydberg_scars}
shows the probability density for the quantum scarred state
whose energy is
closest to $7C_6/R^6$
for each $\rho$.
It is maximal near the
three classical trajectories $B$. 
This signals a stabilization of trajectory $B$,
whose origin is purely quantum since the unstable trajectories
belong to the ergodic region of classical phase space
(see Fig.~\ref{fig:SoS}).

\section{Semiclassical  analysis}

\begin{table}
  \begin{tabular}{|c|c|c|c|c|}
    \hline
    & $T_B^{(\rho)}$ & $S_B^{(\rho)}$ & $\alpha_B^{(\rho)}$ & $k^{(\rho)}$\\
    \hline
    $A_1$ & $T_B/2$ & $S_B/2$ & $\lambda_B T_B/2$& $k$\\
    $A_2$ & $T_B/2$ & $S_B/2$ & $\lambda_B T_B/2$& $k-1/2$\\
    $E$   & $2T_B$  & $S_B$   & $\lambda_B T_B$  & $k+1/2$\\
    \hline
  \end{tabular}
  \caption{\label{tab:traceformulaparams}
    Parameters $T_B^{(\rho)}$, $S_B^{(\rho)}$, $\alpha_B^{(\rho)}$, $k^{(\rho)}$
    for Eq.~\ref{eq:traceformula}, depending on the
    irreducible representation $\rho=A_1$, $A_2$, or $E$.}
\end{table}

\begin{figure}
  \includegraphics[width=\linewidth]
  {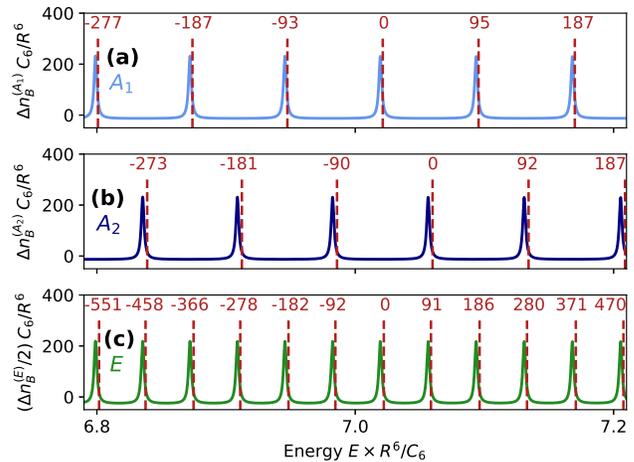}
  \caption{\label{fig:semicl}
    For each irreducible representation $\rho=$
    \textit{(a)} $A_1$, \textit{(b)} $A_2$ and \textit{(c)} $E$,
    the solid curve shows the
    semiclassical contribution $\Delta n_B$
    (Eq.~\ref{eq:traceformula}) to the density of
    states $n$ due to the periodic trajectory $B$,
    as a function of the energy $\epsilon$.
    The dashed vertical lines show the energies of the scarred quantum states,
    which closely match
    the maxima of $\Delta n_B$.
    The integers above them specify the relative state
    indices $\Delta\nu^{A_1}$, $\Delta\nu^{A_2}$, $\Delta\nu^{E}/2$
    with respect to the index $\nu^{(\rho)}$
    of the scarred states in Fig.~\ref{fig:Rydberg_scars}.
  }
\end{figure}

For the majority of the calculated quantum states,
the probability density $|\psi(x,y)|^2$ is unrelated to
the periodic trajectories of type B. Nevertheless,
for  each representation,
we find multiple  scarred quantum states, represented by the
vertical   dashed  lines   in
Fig.~\ref{fig:semicl},
whose energy
spacing   is  approximately   regular.
This is analogous to the tower of scarred
many--body states
with an approximately constant
energy separation
found in a PXP chain  \cite{turner:NatPhys2018}, which
is a recurrent feature in theoretical analyses of weak ergodicity
breaking
\cite{choi:PRL2019,serbyn:NatPhys2021,moudgalya:RepProgPhys2022}.
In the present context, we explain
the series of scarred quantum states semiclassically.
We use Gutzwiller's trace formula
\cite[chap.~17]{gutzwiller:Springer1990}
describing the impact of the classical periodic trajectories on
the quantum density of states $n(\epsilon)$.
We isolate the contribution $\Delta n_B^{(\rho)}$
to $n$ coming from the unstable trajectory $B$,
which depends on the representation $\rho$
\cite{robbins:PRA1989,lauritzen:PRA1991}:
\begin{multline}
  \label{eq:traceformula}
  (\Delta n_B^{(\rho)} \: 2\pi\hbar/T_B^{(\rho)}+1)/\alpha_B^{(\rho)}=\\
    \sum_{k=0}^{\infty}
    \{[S_B^{(\rho)}/\hbar-2\pi(k^{(\rho)}+1/2)]^2+(\alpha_B^{(\rho)}/2)^2\}^{-1}
    \ .
  \end{multline}
The parameters
$T_B^{(\rho)}(\epsilon)$,
$S_B^{(\rho)}(\epsilon)$,
$\alpha_B^{(\rho)}(\epsilon)$, and $k^{(\rho)}$
in Eq.~(\ref{eq:traceformula})
are defined in Table~\ref{tab:traceformulaparams} for each representation.
They are directly related to the classical period $T_B(\epsilon)$
and action $S_B(\epsilon)=\oint \vec{p}\cdot d\vec{x}$ along
one trajectory $B$,
the product $\lambda_B(\epsilon) T_B(\epsilon)$,
and the summation index $k$, respectively.
Figure~\ref{fig:semicl} shows $\Delta n_B^{(\rho)}(\epsilon)$
for each representation.
Its maxima
agree with the energies of the 
scarred states. Hence, the series of scarred states
found in each representation
reflects the multiple resonances in $n(\epsilon)$ due to  
the unstable trajectory $B$. The regularity in their
energy spacing follows from
the resonance maxima being evenly spaced in terms of the classical action,
$S_{B\mathrm{max}}^{(\rho)}/\hbar=2\pi(k^{(\rho)}+1/2)$.

\section{Experimental prospects and outlook}
We consider e.g.\ ${}^{87}\mathrm{Rb}$ atoms in the $50C$
circular Rydberg state \cite{nguyen:PRX2018,brune:PRR2020},
for which $C_6/h=3\,\mathrm{GHz}\,\mathrm{\mu m}^6$. The value $\eta=0.01$
corresponds to $R=7\,\mu\mathrm{m}$. The ring--shaped trap
may be realized optically using Laguerre--Gauss laser beams
and light sheets \cite[\S II.C.2]{amico:RMP2022}.
The energy $\epsilon=7C_6/R^6=h\times 200 \,\mathrm{kHz}$ is within experimental reach
\cite{nguyen:PRX2018}.
For small angular momenta, the centrifugal energy, which is proportional to $(\eta n)^2/3$,
is negligible compared to $\epsilon$.
The position of the atoms may be detected at a given time by turning on a 2D optical
lattice trapping individual Rydberg atoms  \cite{nguyen:PRX2018,raithel:PRL2011},
which freezes the dynamics,
followed by atomic deexcitation and site--resolved ground--state imaging
\cite{scholl:PRXQuantum2022}.

Further investigation will be devoted to the stability of the
quantum scar. Recent experiments \cite{bluvstein:Science2021,su:arXiv2022}
have shown that it may be enhanced
by periodically modulating the parameters.
Depending on the  stabilization mechanism
(see e.g.\ Ref.~\cite{hudomal:PRB2022}
or Ref.~\cite[\S 27]{landau1:ButterworthHeinemann1976}),
this may lead to 
a discrete time crystal \cite{else:ARCMP2020} which is either quantum or classical.

\begin{acknowledgments}
  We acknowledge  stimulating discussions with
  M.~Brune and J.M.~Raimond (LKB, Coll\`ege de France)
  and R.J.~Papoular (IRAMIS, CEA Saclay).
\end{acknowledgments}

\appendix
\section{}

The goal  of this  Appendix  is twofold.
In Section \ref{sec:SM:HenonHeiles}, we
identify  novel   quantum  scars  supported  by   the  H\'enon--Heiles
Hamiltonian,  and  characterize  them  using  the  same  semiclassical
argument as  in the  main text.
In Section \ref{sec:SM:boundaryconditions},
for  each of  the three
irreducible representations of the  group $C_{3v}$, we derive boundary
conditions  defining  quantum  stationary states  within  the  reduced
configuration space.

\subsection{\label{sec:SM:HenonHeiles}
  Quantum scars in the H{\'e}non--Heiles model}
In this Section, we briefly describe our results, analogous to those of the main text,
for the Henon--Heiles Hamiltonian \cite{henon:AstronJ1064}
$H_{\mathrm{HH}}=(p_x^2+p_y^2)/(2m)+V_\mathrm{HH}$, where:
\begin{equation}
  \label{eq:HHenonHeiles_fullunits}
  V_{\mathrm{HH}}=
  m\omega_0^2(x^2+y^2)/2+\alpha(x^2y-y^3/3)
  \ .
\end{equation}
Equation \ref{eq:HHenonHeiles_fullunits} is written  in the
dimensional form of Ref.~\cite[\S 5.6.4]{brack:AddisonWesley1997}
which assumes that the coordinates $x$ and $y$ carry the unit of length.
The quantities $p_x$, $p_y$ are their conjugate momenta, the parameters $m$ and
$\omega_0$ denote a mass and a frequency,
and the coefficient $\alpha$ sets the strength of the cubic term.
If lengths, momenta, energies, and times are expressed in units of
$L_\mathrm{HH}=m\omega_0^2/\alpha$,
$P_\mathrm{HH}=m^2\omega_0^3/\alpha$,
$E_\mathrm{HH}=m^3\omega_0^6/\alpha^2$,
$T_\mathrm{HH}=1/\omega_0$,
the dimensionless form matches that of Ref.~\cite{henon:AstronJ1064}.
As in the main text, in terms of these units,
the classical dynamics is independent of $m$, $\omega_0$ $\alpha$.
As for quantum physics, the classical scaling no longer holds,
and the energy spectra and wavefunctions depend on the dimensionless parameter 
$\eta_{\mathrm{HH}}=\hbar/(L_\mathrm{HH}P_\mathrm{HH})=\hbar\alpha^2/(m^3\omega_0^5)$.
Smaller values of $\eta_{\mathrm{HH}}$ signal deeper quasiclassical behavior.

The H\'enon--Heiles potential is related to our main discussion for two reasons.
First, its symmetry group is $C_{3v}$ \cite{lauritzen:PRA1991},
which is the point group of the system analyzed in the main text.
Second, expanding Eq.~(\ref{eq:pot_Rydberg}) there to third order in $x$ and $y$ near
the equilibrium position $O$ shows that it reduces to
Eq.~\ref{eq:HHenonHeiles_fullunits} in the low--energy limit.

The H\'enon--Heiles Hamiltonian has been extensively studied
(see e.g.\ Ref.~\cite[\S 1.4]{lichtenberg:Springer1992}).
Our goal in revisiting it was twofold.
First, we have calibrated our codes against published results for this potential.
Second, we have identified
quantum scars for the H\'enon--Heiles Hamiltonian which, to the best of our knowledge, are
novel. At the end of the section, we point out the relevance of the H\'enon-Heiles potential
in relation to a broad family of systems, which includes the case of dipolar particles.  

\subsubsection{Calibration}
We have used our codes to reproduce the known classical periodic trajectories of $H_\mathrm{HH}$,
their periods and Lyapunov exponents \cite{davies:Chaos1992},
and its surfaces of section for various energies \cite{gustavson:AstronJ1966}.
We have also recovered the quantum energy levels and wavefunctions,
belonging to all three representations,
in  Refs.~\cite{davis:JChemPhys1979,feit:JComputPhys1982}
for $\eta_\mathrm{HH}=1/80$
and in Ref.~\cite{brack:Chaos1995} for $\eta_\mathrm{HH}=0.06^2$.

\subsubsection{Quantum scars for the H\'enon--Heiles potential}

\begin{figure*}
  \includegraphics[width=\linewidth]
  {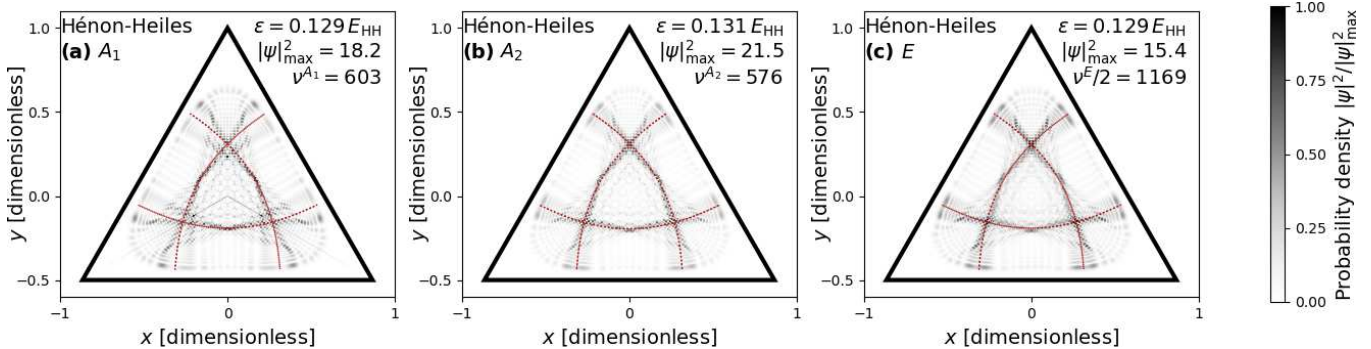}
  \caption{\label{fig:HenonHeiles_scars}
    Probability density $|\psi_\mathrm(x,y)|^2$
    of the scarred quantum eigenstate of the
    H\'enon--Heiles Hamiltonian $H_\mathrm{HH}$ whose
    energy is closest to $0.13E_\mathrm{HH}$
    in each irreducible representation
    \textit{(a)} $A_1$, \textit{(b)} $A_2$, and \textit{(c)} $E$.
    The dashed red lines show
    the three classically unstable periodic trajectories of type
    $B$ for the corresponding
    energy $\epsilon$.
    The 
    densities are maximal near the unstable trajectories,
    signaling the quantum scar.
    The integer $\nu^{(\rho)}$ is the index of the shown quantum state
    in the representation $\rho$.
    (This figure is the analog, for the H\'enon--Heiles potential,
    of Fig.~\ref{fig:Rydberg_scars} in the main text.)
  }
\end{figure*}
We now turn to the lower value $\eta=0.04^2$, so as to consider the deep quasiclassical
regime. We focus on energies $\epsilon\sim 0.13\, E_\mathrm{HH}$: these are large enough
for the ergodic region to occupy a substantial part of phase space \cite{gustavson:AstronJ1966},
while remaining below the threshold energy $E_\mathrm{HH}/6$ above which $H_\mathrm{HH}$
supports trajectories that are not bound \cite{davies:Chaos1992}.
Figure \ref{fig:HenonHeiles_scars} shows the probability density density for the scarred
state with the energy $\epsilon$ closest to $0.13\, E_\mathrm{HH}$ for each representation.
It is maximal near the three trajectories $B$ for the energy $\epsilon$, signaling the scar.

\begin{figure}
  \includegraphics[width=\linewidth]
  {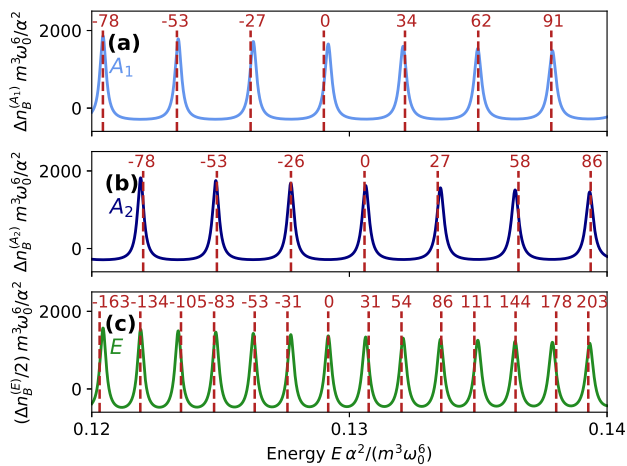}
  \caption{\label{fig:HenonHeiles_semicl}
    For each irreducible representation $\rho=$
    \textit{(a)} $A_1$, \textit{(b)} $A_2$, and \textit{(c)} $E$,
    the solid curve shows the semiclassical contribution
    $\Delta n_B$ to the density of states $n$
    of the H\'enon--Heiles potential
    due to the periodic trajectory $B$,
    as a function of the energy $\epsilon$.
    The dashed vertical lines show the energies of the scarred quantum states,
    which closely match the maxima of $\Delta n_B$.
    The integers above them specify the relative state indices
    $\Delta\nu^{A_1}$, $\Delta\nu^{A_2}$, $\Delta\nu^E/2$ with respect to
    the index $\nu^{(\rho)}$ of the scarred states in
    Fig.~\ref{fig:HenonHeiles_scars}.
    (This figure is the analog, for the H\'enon--Heiles potential, of
    Fig.~\ref{fig:semicl} in the main text.)} 
\end{figure}
In each irreducible representation $\rho=A_1$, $A_2$, and $E$, we find
multiple  scarred quantum  states  for  the H\'enon--Heiles  potential
(vertical  dashed  lines in  Fig.~\ref{fig:HenonHeiles_semicl})  whose
energy spacing  is approximately regular,  in direct analogy  with the
results  of the  main  text.  They may  be  explained  using the  same
semiclassical  argument  relying  on Gutzwiller's  trace  formula.  We
isolate  the  contribution $\Delta  n_B^{(\rho)}$  to  the density  of
states $n$ for  each representation $\rho$ due to
the unstable trajectory $B$.  Both Eq.~4  and Table I in the main text
are applicable  to the H\'enon--Heiles  potential with no  change.  We
have calculated the  required period $T_B$, action  $S_B$ and Lyapunov
exponent $\lambda_B$ characterizing the periodic trajectory $B$ in the
H\'enon--Heiles potential as a function of the energy $\epsilon$ using
our  codes. Figure~\ref{fig:HenonHeiles_semicl}  shows $\Delta n_B^{(\rho)}$
for  each representation  $\rho$.  Just  like  in the  main text,  its
maxima coincide with  the energies of the scarred  states.  Hence, the
same  conclusion holds,  and we  may ascribe  the regularity  in their
energy spacing to  the resonance maxima being equally  spaced in terms
of the classical action $S_B$.

\subsubsection{Generality of the H\'enon--Heiles potential}

The potential $V_\mathrm{HH}$ combines a 2D isotropic harmonic trap
with a two--variable cubic polynomial function.
Hence, it may be seen as the simplest possible
2D potential exhibiting $C_{3v}$ symmetry. The three-body Hamiltonian given by Eq.~1
in the main text reduces to it near one of its (equivalent) minima
for the repulsive pair--wise interaction
$v(d_{ij})=a\,d_{ij}^{-\alpha}$ regardless of the power law exponent $\alpha>0$.
The presence of quantum scars in the H\'enon--Heiles model leads us to expect similar
scars in all of these systems.
In particular, the dipole--dipole interaction \cite{baranov:ChemRev2012}
in the case where all three dipole moments
are  polarized perpendicular to the plane,
corresponding to $\alpha=3$, 
is expected to yield the same phenomena.

\subsection{ \label{sec:SM:boundaryconditions}
  Boundary conditions defining a basis of
  quantum stationary states}

In this Section, we exploit the spatial symmetries of the point group $C_{3v}$
and time--reversal symmetry to state boundary conditions uniquely defining
a basis of quantum stationary states.
We state our reasoning in terms of the system considered in
the main text, but it applies without change to the H\'enon--Heiles
Hamiltonian discussed in Sec.~\ref{sec:SM:HenonHeiles} above.

We  expect the  quantum  states scarred  by  the classically  unstable
periodic  trajectory $B$  to exhibit  an enhanced  probability density
along all three  trajectories $B$ at a given energy  (red dotted lines
in Figs.~\ref{fig:circle_triangle}a
and \ref{fig:Rydberg_scars}a--c in the main text for the system discussed there,
and in  Figs.~\ref{fig:HenonHeiles_scars}a--c in the  present
Appendix 
for  the  H\'enon--Heiles  model).  Hence,  the  probability
density  for  the  scarred  states is  expected  to  exhibit  $C_{3v}$
symmetry. Therefore, we construct a basis of quantum stationary states
whose corresponding density profiles all exhibit this symmetry. This
property is not automatically satisfied and requires choosing appropriate
basis functions. For example, Figs.~7a and 7b in
Ref.~\cite{feit:JComputPhys1982} show probability densities corresponding
to eigenstates of the H\'enon--Heiles model which do not exhibit $C_{3v}$ symmetry
despite the fact that the Hamiltonian does, see Sec.~\ref{sec:SM:HenonHeiles} above.

The group $C_{3v}$ admits 3 irreducible representations,
$\rho=A_1$, $A_2$, and $E$
\cite[\S 95]{landau3:ButterworthHeinemann1977}. Representations $A_1$ and $A_2$
are 1D, whereas representation $E$ is 2D.
For each representation, we shall formulate
a boundary condition defining basis functions belonging to it.
All wavefunctions $\psi(\vec{r})$ are normalized according to
$\iint_{ABC} d^2r|\psi(\vec{r})|^2=1$, the integral being taken over the
triangle $ABC$.

\subsubsection{\label{sec:irreps_A1_A2}
  One--dimensional representations $A_1$ and $A_2$}
\begin{figure}
  \includegraphics[width=\linewidth]
  {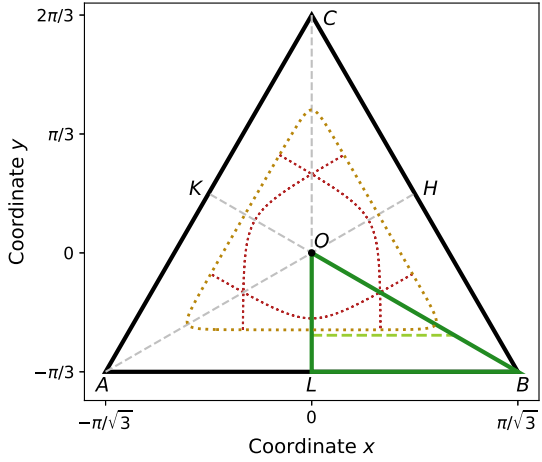}
  \caption{\label{fig:boundaryconditions}
    The black triangle $ABC$ is the classical configuration space for the
    Hamiltonian $H_\mathrm{2D}$ of the main text. The smaller green triangle
    $OLB$ is the reduced configuration space
    within which we solve for the quantum stationary states.
    The classically accessible region, limited by the dotted golden line,
    is shown for the energy $\epsilon=7C_6/R^6$.
    We enforce the boundary condition $\psi=0$ on the quantum  wavefunctions
    along the horizontal
    dashed green line.
    The three classical periodic trajectories $B$ (dotted red lines)
    are also shown.
    }
\end{figure}
We first consider a 1D representation $\rho=A_1$ or $A_2$.
Let $\psi$ be an eigenstate of $H_\mathrm{2D}$ for the energy $\epsilon$
transforming according to $\rho$.
We call $\mathcal{S}_1$, $\mathcal{S}_2$, $\mathcal{S}_3=\mathcal{S}_{\Delta}$
the reflections about 
$(AH)$, $(BK)$, $(CL)$ in the $(x,y)$ plane
(see Fig.~\ref{fig:boundaryconditions}).
The wavefunction $\psi(\mathcal{S}_i\vec{r})$ is also an eigenstate of $H_\mathrm{2D}$
for the same energy $\epsilon$. Because $\rho$ is 1D, 
$\psi(\mathcal{S}_i\vec{r})=\chi_i \psi(\vec{r})$ for some complex number $\chi_i$.
The reflections $\mathcal{S}_i$
satisfy $\mathcal{S}_i^2=1$, so that $\chi_i=\pm 1$.
They also satisfy 
$\mathcal{S}_2\mathcal{S}_1=\mathcal{S}_3\mathcal{S}_2=\mathcal{R}$,
with $\mathcal{R}$ being the rotation of angle $2\pi/3$
about the point $O$. The transformation
$\mathcal{R}^3=1$,
so that $(\chi_1\chi_2)^3=(\chi_2\chi_3)^3=1$. Hence,
$\chi_1=\chi_2=\chi_3=\pm 1$.

The case $\chi_1=\chi_2=\chi_3=1$  leads to
$\rho(\mathcal{R})=\rho(\mathcal{S}_i)=1$, so that $\rho=A_1$
\cite[\S 95, Table 7]{landau3:ButterworthHeinemann1977}.
Then, $\psi(S_i\vec{r})=+\psi(\vec{r})$, leading to the boundary condition
$\partial_n\psi=0$ along the sides $[LO]$ and $[OB]$ of the green triangle $OBL$
in Fig.~\ref{fig:boundaryconditions}. Combined with the condition $\psi=0$
along the side $[LB]$ derived in the main text, it defines a basis
of wavefunctions $\psi$ for Representation $A_1$.

The case $\chi_1=\chi_2=\chi_3=-1$ leads to
$\rho(\mathcal{R})=1$ and $\rho(\mathcal{S}_i)=-1$, so that $\rho=A_2$.
Then, $\psi(\mathcal{S}_i\vec{r})=-\psi(\vec{r})$, leading to the condition $\psi=0$
along the sides $[LO]$ and $[OB]$. Hence, imposing the Dirichlet boundary condition
on the three edges of the triangle $OBL$ defines a basis of  wavefunctions $\psi$ for
Representation $A_2$.

The energy levels transforming according to the 1D representations $A_1$ and $A_2$
are non--degenerate, hence, the time--reversal invariance
of $H_\mathrm{2D}$ allows us to choose all basis wavefunctions $\psi(\vec{r})$
to be real \cite[\S 18]{landau3:ButterworthHeinemann1977}.
Furthermore,
$\psi(\vec{r})=\psi(\mathcal{R}^{-1}\vec{r})=\chi_i \psi(\mathcal{S}_i\vec{r})$
differ by a sign at most. Hence, the corresponding probability densities coincide,
and $|\psi(\vec{r})|^2$ does exhibit $C_{3v}$ symmetry.

\subsubsection{\label{sec:irrep_E}
  Two--dimensional representation $E$}
We now turn to the 2D representation $\rho=E$. Let $\epsilon$ be
a twice--degenerate energy level
of $H_\mathrm{2D}$. The corresponding eigenspace 
is spanned by two complex wavefunctions, $\phi_+$ and $\phi_-$ which transform
according to $\rho$:
\begin{subequations}
  \label{eq:Etransforms_phi_RS}
  \begin{align}
    \label{eq:Etransforms_phi_RS_R}
    \phi_\pm(\mathcal{R}^{-1}\vec{r}) &=  e^{\pm i2\pi/3}  \phi_\pm(\vec{r})
                                        \ ,
    \\
    \label{eq:Etransforms_phi_RS_S}
    \phi_\pm(\mathcal{S}_\Delta \vec{r}) &=  \phi_\mp (\vec{r})
                                           \ ,
  \end{align}
\end{subequations}
where the transformations $\mathcal{R}$ and $\mathcal{S}_\Delta$ are defined as
in Sec.~\ref{sec:irreps_A1_A2} above and the main text.

The time--reversal invariance \cite[\S 18]{landau3:ButterworthHeinemann1977}
of $H_\mathrm{2D}$ entails that the complex--conjugate
wavefunctions $\phi_+^*(\vec{r})$ and $\phi_-^*(\vec{r})$ are also eigenstates of $H_\mathrm{2D}$
with the same energy $\epsilon$. Complex--conjugating Eqs.~(\ref{eq:Etransforms_phi_RS_R}),
accounting for normalization, and writing $(\phi_+^*)^*=\phi_+$
lead to $\phi_\pm^*=e^{i\alpha}\phi_\mp$,
where $e^{i\alpha}$ is a complex number of modulus 1. Introducing the new basis
wavefunctions $\psi_+(\vec{r})=e^{i\alpha/2}\phi_+(\vec{r})$ and
$\psi_-(\vec{r})=\psi_+^*(\vec{r})$,
Eqs.~(\ref{eq:Etransforms_phi_RS}) reduce to 2 conditions on $\psi_+$:
\begin{subequations}
  \label{eq:Etransforms_psi_RS}
  \begin{align}
    \label{eq:Etransforms_psi_RS_R}
    \psi_+(\mathcal{R}^{-1}\vec{r}) &=  e^{i2\pi/3}  \psi_+(\vec{r})
                                        \ ,
    \\
    \label{eq:Etransforms_psi_RS_S}
    \psi_+(\mathcal{S}_\Delta \vec{r}) &=  \psi_+^* (\vec{r})
                                           \ .
  \end{align}
\end{subequations}
The probability densities
$|\psi_+(\vec{r})|^2=|\psi_+(\mathcal{R}^{-1}\vec{r})|^2=|\psi_+(\mathcal{S}_\Delta\vec{r})|^2$
coincide. Hence, $|\psi_+(\vec{r})|^2$ exhibits $C_{3v}$ symmetry: this is the probability
density plotted in Figs.~4a--c of the main text (three Rydberg atoms)
and Figs.~\ref{fig:HenonHeiles_scars}a--c (H\'enon--Heiles model).

We seek $\psi_+(\vec{r})$ in the following form, which is more amenable to numerical computation:
\begin{equation}
  \label{eq:psiplus_u1_u2}
  \psi_+(\vec{r})=(x-iy)(u_1(\vec{r})+iu_2(\vec{r}))
  \ ,
\end{equation}
where  $u_1$  and $u_2$  are  two  real functions  satisfying  coupled
Schr\"odinger equations.  In  Eq.~(\ref{eq:psiplus_u1_u2}), the factor
$(x-iy)$ accounts for the fact  that $\psi_+(\vec{0})=0$, like for the
stationary  states of  the 2D  isotropic harmonic  oscillator carrying
angular   momentum  \cite[\S   112]{landau3:ButterworthHeinemann1977}.
Equations  \ref{eq:Etransforms_psi_RS} yield  the boundary  conditions
$u_1=0$,  $\partial_n  u_2=0$  along  both  $[LO]$  and  $[OB]$  (see
Fig.~\ref{fig:boundaryconditions}).    Combined  with   the  condition
$\psi=0$ along $[LB]$ derived in the main text, they define a basis of
stationary  states related  to  Representation $E$. For  each of  the
twice--degenerate  energy   levels,  $\psi_+(\vec{r})$  is   given  by
Eq.~(\ref{eq:psiplus_u1_u2})   and  the   second  basis   function  is
$\psi_+^*(\vec{r})$.

\subsubsection{Spatial extent of the wavefunctions}
For  a  given energy  level  $\epsilon$,  the  spatial extent  of  the
stationary   states   defined  in   Secs.~\ref{sec:irreps_A1_A2}   and
\ref{sec:irrep_E}  barely  exceeds  the classically  accessible  region
(limited       by      the       dotted      golden       line      in
Fig.~\ref{fig:boundaryconditions} for  the Hamiltonian $H_\mathrm{2D}$
of the main text and $\epsilon=7C_6/R^6$).  Therefore, we restrict the
region within  which we solve for  the wavefunctions to a  part of the
triangle $OLB$ which slightly exceeds  this region. In other words, we
enforce the  condition $\psi=0$ not  on $[LB]$, but on  the horizontal
dashed line in Fig.~\ref{fig:boundaryconditions}.

\subsubsection{Indices of the quantum states}
We order the  quantum states pertaining to a  given irreducible representation
$\rho$  by increasing  energies. This  gives rise  to the  state index
$\nu^{(\rho)}$  appearing in  Figs.~\ref{fig:Rydberg_scars}
and  \ref{fig:semicl} in  the  main text,  and
Figs.~\ref{fig:HenonHeiles_scars} and  \ref{fig:HenonHeiles_semicl} in
the  present Appendix. 
The  irreducible representations
$A_1$  and  $A_2$  have  dimension  1,  so  that,  barring  accidental
degeneracies, the corresponding energy levels are non--degenerate.  By
contrast, the irreducible representation  $E$ has dimension 2, meaning
that each energy  level is twice degenerate.  For this representation,
we consistently indicate  one half of the state  index, $\nu^{(E)}/2$, and
one half of the density of states contribution $\Delta n_B^{(E)}/2$.

The relative level indices $\Delta\nu^{(\rho)}$ given in
Fig.~\ref{fig:semicl} of the
main   text   and   in   Fig.~\ref{fig:HenonHeiles_semicl}   of   this
Appendix 
are   exact.    The   level    indices   of
Fig.~\ref{fig:HenonHeiles_scars},   concerning   the   H\'enon--Heiles
model, are also exact.  We  obtain approximations to the level indices
of  Fig.~\ref{fig:Rydberg_scars}
in  the main  text, concerning  three Rydberg  atoms moving
along a circle, using the  semiclassical appproximation to the density
of  states, accounting  for the  role of  discrete spatial  symmetries
\cite{lauritzen:AnnPhys1995}.

%

\end{document}